\pdfoutput=1
\documentclass[a4paper]{article}
\usepackage{INTERSPEECH2020}

\usepackage{amssymb}
\usepackage{multirow} 
\graphicspath{{grf/}}

\usepackage{algorithm}
\usepackage{algorithmic}

\usepackage{epstopdf}

\newcommand{\calC}{\ensuremath{\mathcal{C}}}

\newcommand{\calH}{\ensuremath{\mathcal{H}}}

\newcommand{\calT}{\ensuremath{\mathcal{T}}}

\newcommand{\caja}[4][1]{{%
    \renewcommand{\arraystretch}{#1}%
    \begin{tabular}[#2]{@{}#3@{}}%
      #4%
    \end{tabular}%
    }}

{%
\begin{list}{#1}{
\vspace{-\topsep}
\vspace{-\partopsep}
\setlength{\itemindent}{0cm}
\setlength{\rightmargin}{0cm}
\setlength{\listparindent}{0cm}
\settowidth{\labelwidth}{#1}
\setlength{\leftmargin}{\labelwidth}
\addtolength{\leftmargin}{\labelsep}
\setlength{\itemsep}{0cm}
}%
}%
{%
\end{list}
\vspace{-\topsep}
\vspace{-\partopsep}
}

{\begin{enumerate}%
}%
{\end{enumerate}}

\usepackage{color}
\usepackage{colortbl}
\definecolor{light-gray}{gray}{0.8}

\title{An investigation of phone-based subword units for end-to-end speech recognition}
\name{Weiran Wang $\;$ Guangsen Wang $\;$ Aadyot Bhatnagar $\;$ Yingbo Zhou $\;$ Caiming Xiong $\;$ Richard Socher}
\address{Salesforce Research \\ 575 High St, Palo Alto, CA 94301, USA}
\email{\{weiran.wang, guangsen.wang, abhatnagar, yingbo.zhou, cxiong, rsocher\}@salesforce.com}

\begin{document}
\maketitle

\begin{abstract}
Phones and their context-dependent variants have been the standard modeling units for conventional speech recognition systems, while characters and subwords have demonstrated their effectiveness for end-to-end recognition systems.
We investigate the use of phone-based subwords, in particular, byte pair encoder (BPE), as modeling units for end-to-end speech recognition.
In addition, we also developed multi-level language model-based decoding algorithms based on a pronunciation dictionary.
Besides the use of the lexicon, which is easily available, our system avoids the need of additional expert knowledge or processing steps from conventional systems.
Experimental results show that phone-based BPEs tend to yield more accurate recognition systems than the character-based counterpart.
In addition, further improvement can be obtained with a novel one-pass joint beam search decoder, which efficiently combines phone- and character-based BPE systems.
For Switchboard, our phone-based BPE system achieves 6.8\%/14.4\% word error rate (WER) on the Switchboard/CallHome portion of the test set while joint decoding achieves 6.3\%/13.3\% WER. 
On Fisher + Switchboard, joint decoding leads to 4.9\%/9.5\% WER, setting new milestones for telephony speech recognition.
\end{abstract}
\section{Introduction}
\label{s:intro}
\vspace*{-1ex}

For English speech recognition, phones and the context-dependent variants have long been the standard modeling units for conventional automatic speech recognition (ASR) systems~\cite{Young_99a,Povey_11a}. However, there has been a surge of interest in using character and character-based subwords, 
such as byte pair encoding (BPE,~\cite{Sennric_16a}) and word-pieces~\cite{SchustNakajim12a}, in modern end-to-end systems~\cite{GravesJaitly14a,Miao_15a,Sak_15a,Amodei_16a,Collob_16a,Chan_16a,Watanab_18a,Zeyer_18a,He_18a,Wang_19a}. 
The advantages of character and subwords mainly lie in the simplicity. 
They not only enable straightforward open-vocabulary recognition, but also avoid the need of domain knowledge such as pronunciation dictionary and phonetic questions in state tying~\cite{Young_94b} for context-dependent phones, and additional processing steps such as hidden Markov models and Gaussian mixture models training for phone state modeling. 
On the other hand, phones have tighter correspondence to audio than characters, and often leads to higher recognition accuracy. As a concrete example, for the Switchboard corpus, it takes more modeling effort for the char CTC system to match performance of the phone CTC system, as noted by~\cite{Miao_16a} and~\cite{Zweig_17a}. This motivates the question of whether phone-based subwords would be more effective for ASR.

In this work, we investigate the use of phone-based BPEs in the context of end-to-end speech recognition. We use a pronunciation dictionary to convert transcription into phone sequences while maintaining the word boundaries, and extract BPEs by gradually merging frequent pairs of phones or phone sequences, as is done for character BPEs.\footnote{We use implementation of \cite{Kudo18a} for training and inference of BPEs.}
Intuitively, similar to the context-dependent phones, phone BPEs shall capture correlations between contiguous phones. On the other hand, it allows a trade-off between the size of modeling units and output 
length, which influences both the performance and efficiency of end-to-end systems. We train the acoustic model using the phone BPE targets as usual, with the hybrid CTC/attention loss~\cite{Watanab_17a}. 

At decoding time, we use the pronunciation dictionary again to convert decoded phone BPE sequence back into words, with a novel multi-level RNN language model (LM).
To take advantage of the complementarity of both phone and character BPE systems, we develop a one-pass joint beam search decoder that efficiently combines the two.
Our experimental results on both the Wall Street Journal and Switchboard show that the phone BPE systems tend to outperform the character-based counterpart in accuracy, and joint decoding with the two leads to significant improvement. Specifically, our phone BPE system achieves 6.8\%/14.4\% word error rate (WER) on the SWBD/CALLHM portion of the test set while joint decoding achieves 6.3\%/13.3\% WER. This, to the best of our knowledge, sets the new state-of-the-art on Switchboard 300 hours. 
By adding the Fisher corpora into training, we obtain best single systems and joint decoding yields 4.9\% WER on the SWBD portion, setting a new milestone for this setup.

We would like to emphasize that, besides the use of pronunciation dictionary for subword extraction and decoding\footnote{We pick one pronunciation for each word if the word has multiple pronunciations in the dictionary.}, our method avoids the extra processing steps from conventional systems. On the other hand, large collections of pronunciations are readily accessible~\cite{cmudict}, and pronunciation of out-of-collection words can be constructed with grapheme-to-phoneme methods~\cite{BisaniNey08a,Rao_15b,ToshniwLivesc16a,Yolchuy_19a}, and therefore our approach maintains the simplicity of end-to-end methods.


\section{Multi-level LM for decoding with BPEs}

\label{s:decode}
\vspace*{-1ex}

Compared to decoding with character BPEs which just needs to output the highest scoring sequence of modeling units, possibly without additional language models (subword-level or word-level), there are a few challenges for decoding with phone BPEs.
First, it is necessary to use a pronunciation dictionary to convert the decoded phone sequence into a word sequence. Second, unlike in the character case where the spelling uniquely determines a word, different words can have the same pronunciations for the case of phones (i.e., homophones) and therefore a word LM is helpful for distinguishing them.

We develop a multi-level LM that combines scores from both a subword LM and a word LM, and use it in beam search by shallow fusion~\cite{Gulceh_15a}. Intuitively, the method is similar to that proposed in~\cite{Hori_17a}, which combines character LM and word LM. We build a prefix tree storing the pronunciation of words in the dictionary, and as we move down the tree from the root according to the hypothesized subwords and accumulating subword LM scores from each step, we may come across tree nodes that contain words, whose pronunciations match the sequences of subwords on the paths stemming from root. At that time, we may decide to output the words and replace accumulated subword LM scores with word LM scores, and subsequently move back to the root. We highlight challenges not present in~\cite{Hori_17a}:
\begin{itemize}
\item To build the prefix tree, we need to first decompose the pronunciation of each word (i.e., a phone sequence) into a phone BPE sequence, using the BPEs extracted from transcription. This decomposition is greedy (utilizing large subwords as much as possible) and deterministic.
\item In~\cite{Hori_17a}, the modeling units are characters, and one can determine the completion of a word when the word boundary '\_' is proposed.
In the BPE case, however, many subwords contain '\_' as the first symbol.
Therefore, all these subwords indicate both the word boundary and a new word is started,  \emph{simultaneously}.
\item Due to the issue of homophones, a node in the prefix tree may contain multiple words. If the word boundary is met at such a node, we have to output multiple word hypotheses. As a result, one decoding beam branches into multiple beams, which have the same subword LM state but different word LM states. 
\end{itemize}

In Algorithm~\ref{alg:decode-lm}, we detail the forwarding function of our multi-level LM $\mathtt{M}$, constructed from a subword LM $\mathtt{S}$ and a word LM $\mathtt{W}$. This function $\mathtt{forward(state,\,s)}$ takes as input the current state $\mathtt{state}$ and a subword $\mathtt{s}$, and returns the updated state 
after accepting $\mathtt{s}$, the vector of look-ahead scores $\mathtt{la\_{}scores}$ for all subwords, and word outputs if the word boundary is met (output is set to special token $\mathtt{<\!\!incomplete\!\!>}$ otherwise). $\mathtt{la\_{}scores}$ are computed based on scores from $\mathtt{S}$ and $\mathtt{W}$ with a relative weight $\alpha$ ($\mathtt{S}$ is not activated if $\alpha=0$), and they are combined with the acoustic model scores for evaluating partial hypotheses.
The $\mathtt{state}$ for multi-level LM is a 6-tuple \\
\hspace*{2em} $\mathtt{ (Sstate,\, Slogp,\, Wstate,\, Wlogp,\, node,\, accum) }$ \\
\noindent containing the state and associated log-probabilities (for subwords) from $\mathtt{S}$, the state and associated log-probabilities (for words) from $\mathtt{W}$, the position in the prefix tree $\mathtt{T}$, accumulated subword score since the last output word. To start decoding, we initialize states and log-probabilities by accepting the start of sentence token $\mathtt{<\!\!sos\!\!>}$, and set $\mathtt{node}$ to the root of $\mathtt{T}$.

\begin{algorithm}[th!]
\caption{The forwarding function of multi-level RNNLM.}
\label{alg:decode-lm}
  \renewcommand{\algorithmicrequire}{\textbf{Input:}}
  \renewcommand{\algorithmicensure}{\textbf{Output:}}
  \hrule
  \begin{algorithmic}
    \REQUIRE subword $\mathtt{s}$, previous state $\mathtt{state}$. $\mathtt{Wlogp(w)}$ gives the score at position $\mathtt{w}$ of vector $\mathtt{Wlogp}$. Function $\mathtt{node.getWords()}$ returns the list of complete words associated with $\mathtt{node}$ of prefix tree $\mathtt{T}$, $\mathtt{node.getTokens()}$ returns the list of subwords branching out from $\mathtt{node}$, and $\mathtt{node.branch(s)}$ returns the new node after accepting $\mathtt{s}$ at $\mathtt{node}$. $\alpha$ is used for weighing scores of $\mathtt{S}$ versus $\mathtt{M}$. \\[-1ex]\hrulefill
    \STATE $\mathtt{(Sstate,\, Slogp,\, Wstate,\, Wlogp,\, node,\, accum) \leftarrow state}$
    \IF{$\mathtt{s.startswith('\_')}$ and $\mathtt{(not\; node==root)}$}
    \STATE \# \emph{Word boundary is met, inter-word transition}
    \IF{$\mathtt{node.getWords()}$ is not empty}
    \STATE $\mathtt{ wordlist \leftarrow node.getWords() }$
    \ELSE 
    \STATE $\mathtt{ wordlist \leftarrow [<\!\!unk\!\!>] }$
    \ENDIF
    \STATE $\mathtt{ output \leftarrow [\,]}$ \hspace{1em} (\emph{empty list})
    \FOR{$\mathtt{w}$ in $\mathtt{wordlist}$}
    \IF{$\mathtt{ w==<\!\!unk\!\!> }$}
    \STATE $\mathtt{ adjust \leftarrow Wlogp(<\!\!unk\!\!>) + oov\_{}penalty }$
    \ELSE
    \STATE $\mathtt{ adjust \leftarrow Wlogp(w) - accum }$
    \ENDIF
    \STATE \# \emph{Update word LM state}
    \STATE $\mathtt{ (Wstate\_{}new,\, Wlogp\_{}new) \leftarrow W.forward(Wstate,\, w) }$
    \STATE $\hspace*{-1em}\mathtt{ acum\_{}new \leftarrow \alpha \cdot Slogp(s) },\quad \mathtt{ node\_{}new \leftarrow root.branch(s) }$
    \STATE \# \emph{Update subword LM state}
    \STATE $\mathtt{ (Sstate\_{}new,\, Slogp\_{}new) \leftarrow S.forward(Sstate,\, s) }$
    \STATE $\mathtt{ la\_{}scores\_{}new \leftarrow adjust + \alpha \cdot Slogp\_{}new  }$
    \STATE $\mathtt{ state\_{}new \leftarrow (Sstate\_{}new,\, Slogp\_{}new,\, Wstate\_{}new,\,}$
    \STATE $\qquad\qquad\qquad\quad \mathtt{ Wlogp\_{}new,\, node\_{}new,\, accum\_{}new)}$
    \STATE $\mathtt{ output.append((state\_{}new,\, la\_{}scores\_{}new,\, w))}$
    \ENDFOR
    \RETURN $\mathtt{output}$
    \ELSE
    \STATE \# \emph{Intra-word transition, no word output}
    \STATE $\mathtt{w \leftarrow <\!\!incomplete\!\!>}$
    \IF{$\mathtt{s}$ in $\mathtt{node.getTokens()}$}
    \STATE $\mathtt{ node \leftarrow node.branch(s) }$
    \STATE $\mathtt{ accum \leftarrow accum + \alpha \cdot Slogp(s) }$
    \STATE $\mathtt{ (Sstate,\, Slogp) \leftarrow S.forward(Sstate,\, s) }$
    \STATE  $\mathtt{ la\_{}scores \leftarrow \alpha \cdot Slogp }$
    \ELSE 
    \STATE $\mathtt{la\_{}scores \leftarrow } - \mathbf{\infty}$ \hspace{1em} (\emph{vector of all $-\infty$'s})
    \ENDIF
    \STATE $\mathtt{ state \leftarrow (Sstate,\, Slogp,\, Wstate,\, Wlogp,\, node,\, accum) }$
    \RETURN $\mathtt{[(state,\, la\_{}scores,\, w)]}$
    \ENDIF
  \end{algorithmic}
\end{algorithm}

\section{Joint one-pass decoding of BPE systems}
\label{s:joint}
\vspace*{-1ex}

The multi-level LM in Section~\ref{s:decode} applies to both phone and character BPE systems. Since the two types of units can be complementary to each other (capturing different aspects of the language), to take advantage of this, we develop a one-pass beam search algorithm utilizing both systems. Note that, while ideas of (hierarchically) combining phone and character labels have been explored for acoustic model training in end-to-end systems (e.g., ~\cite{SakRao17a,Toshniw_17a,Rao_17a,Yu_18a}), our approach of combining different units at decoding time is orthogonal to them.
In~\cite{Xu_19a}, the authors used the pronunciation dictionary to assist the extraction of character-based subwords, sharing the intuition that phones and characters are complementary. However, in the end they train only one subword system, and the subword extraction requires a non-trivial step of aligning phones and characters. 

Our main idea is to use the phone BPE system, denoted as $\mathtt{Model1}$, to propose subwords. After seeing a word boundary, we decompose the word into a character BPE sequence, and run the character BPE system, denoted as $\mathtt{Model2}$, to accept the sequence. Scores from both systems are then linearly combined up to the word boundary. 
In other words, the phone BPE system leads the decoding process and the character BPE system verifies its hypotheses; the two systems synchronize at each word boundary. In such a way, we incorporate the evidence from $\mathtt{Model2}$ as early as possible to adjust scores of word hypotheses. Compared to second-pass rescoring, our one-pass approach avoids generating large amount of hypotheses by $\mathtt{Model1}$. 

To simplify presentation, we divide each BPE system into acoustic model (AM) and language model (LM). The AM refers to the model trained with end-to-end objectives, e.g., the hybrid CTC/attention model~\cite{Watanab_17a} in our case (although the decoder implicitly models the language of subwords). The AM also provides a scoring function which computes the score of the next subword given acoustic inputs and previously decoded subwords (in our case, the score is a linear combination of log-probabilities from the attention decoder and CTC prefix score). The LM refers to the one described in previous section with its components (subword and word LMs) trained separately on text data, and provides the $\mathtt{forward()}$ function which computes the score of next subword given previously decoded results.

Our algorithm maintains a set of decoding beams (hypotheses), each of which is an 8-tuple\\
\hspace*{3em} $\mathtt{(score,\, ws,\, sc1,\, ys1,\, st1,\, sc2,\, ys2, \, st2)}$\\
\noindent containing the final score of the beam (for pruning), the word hypothesis, followed by the score ($\mathtt{sc}$), output subword sequence ($\mathtt{ys}$), and multi-level LM state ($\mathtt{st}$) from $\mathtt{Model1}$ and $\mathtt{Model2}$ respectively.
The detailed procedure is given in Algorithm~\ref{alg:decode-joint}. We use the parameter $\beta$ for combining the LM score with the AM score within each system, and $\gamma \in [0,1]$ for combining scores from both systems, as shown in (*). We use the end detection method of~\cite{Watanab_17a} for terminating the algorithm, when longer hypotheses consistently yield much lower scores than the best finished hypothesis (ending with $<\!\!eos\!\!>$).

In our algorithm, for each beam we run $\mathtt{Model1}$ once through the phone BPE sequence, run $\mathtt{Model2}$ once through the corresponding character BPE sequence; $\mathtt{Model2}$ does not propose additional hypotheses but simply follows $\mathtt{Model1}$. Therefore, the time complexity of Algorithm~\ref{alg:decode-joint} is roughly the sum of that of individual systems for the same beam size.

\begin{algorithm}[th!]
\caption{Beam search algorithm for joint BPE system.}

\label{alg:decode-joint}
  \renewcommand{\algorithmicrequire}{\textbf{Input:}}
  \renewcommand{\algorithmicensure}{\textbf{Output:}}
  \hrule
  \begin{algorithmic}
\REQUIRE Input $\mathtt{x}$, trained models, and parameters $(\beta,\,\gamma)$. 
$\mathtt{top(sc, bs)}$ return the list of $\mathtt{(score,\, subword)}$-tuples of the $\mathtt{bs}$ highest values in vector $\mathtt{sc}$. $\mathtt{prune(\calH,\, bs)}$ returns the $\mathtt{bs}$ highest scoring beams (in the $\mathtt{score}$ field) from $\calH$.  $\mathtt{finish(beam)}$ forwards both systems to accept $\mathtt{<\!\!eos\!\!>}$ and the final output words), and compute final score as in $(*)$. \\[-1ex]\hrulefill
    \STATE $\mathtt{\calH \leftarrow [(0.0,\, [<\!\!sos\!\!>],\, 0.0,\, [<\!\!sos\!\!>],\, init\_{}st1,\, }$
    \STATE $\mathtt{\qquad\qquad\qquad\qquad\quad 0.0,\, [<\!\!sos\!\!>],\, init\_{}st2)]}$
    \STATE $\mathtt{\calC \leftarrow [\,]}$ \hspace*{1em} (\emph{set of completed beams})
    \WHILE {$\mathtt{end\_{}detection(\calC)==false}$}
    \STATE $\calT \leftarrow [\,]$
    \FOR{ $\mathtt{beam\; \text{in}\;\calH}$}
    \STATE $\mathtt{ (score,\, ws,\, sc1,\, ys1,\, st1, sc2,\, ys2,\, st2) \leftarrow beam }$
    \STATE $\mathtt{ lm1\_{}output \leftarrow LM1.forward(st1,\, ys1[-1]) }$
    \FOR{$\mathtt{(st1\_{}n,\, la1,\, w)\; \text{in}\; lm1\_{}output}$}
    \STATE $\mathtt{ yscores1 \leftarrow AM1.score(x, ys1) + \beta \cdot la1}$
    \STATE $\mathtt{sc2\_{}n \leftarrow sc2,\quad ys2\_{}n \leftarrow ys2,\quad st2\_{}n \leftarrow st2}$
    \IF{$\mathtt{not\; w==<\!\!incomplete\!\!>}$}
    \STATE \hspace{-1em} \# \emph{Word boundary, forward $\mathtt{Model2}$}
    \STATE \hspace{-1em} $\mathtt{\Delta \leftarrow spm\_{}encode(w)}$  \hspace*{0.2em} (\emph{decomp. $w$ into char BPEs})
    \FOR{$\mathtt{y\;\text{in}\;}\Delta$}
    \STATE \hspace*{-1em} $\mathtt{ (st2\_{}n,\, la2,\, \_)\  \leftarrow LM2.forward(st2\_{}n,\, ys2\_{}n[-1])}$
    \STATE \hspace*{-1em} $\mathtt{ sc2\_{}n \leftarrow sc2\_{}n + AM2.score(x, ys2\_{}n)(y) + \beta \cdot la2(y) }$
    \STATE \hspace*{-1em} $\mathtt{ ys2\_{}n.append(y)}$
    \ENDFOR
    \ENDIF
    \FOR{$\mathtt{(c,\, y)\;\text{in}\;top(yscores1,\, beamsize)}$}
    \STATE $\mathtt{ws\_{}n \leftarrow ws,\; sc1\_{}n \leftarrow sc1 + c,\; ys1\_{}n.append(y) }$
    \IF{$\mathtt{not\; w==<\!\!incomplete\!\!>}$}
    \STATE \# \emph{Incorporate $\mathtt{Model2}$ score at word boundary}
    \STATE $\mathtt{ score\_{}n \leftarrow (1-\gamma) \cdot sc1 + \gamma \cdot sc2\_{}n + c } \quad (*)$
    \STATE $\mathtt{ws\_{}n.append(w)}$
    \ELSE
    \STATE \# \emph{Otherwise update score with $\mathtt{Model1}$ only}
    \STATE $\mathtt{ score\_{}n \leftarrow score + c }$
    \ENDIF
    \STATE $\mathtt{\calT.append( (score\_{}n,\, ws\_{}n,\, sc1\_{}n,\, ys1\_{}n,\,  }$
    \STATE $\mathtt{\qquad\qquad\qquad st1\_{}n,\,sc2\_{}n,\, ys2\_{}n,\,  st2\_{}n))}$
    \ENDFOR
    \ENDFOR
    \ENDFOR
    \STATE $\mathtt{\calH \leftarrow prune(\calT,\, beamsize)}$
    \FOR{$\mathtt{beam\; \text{in}\; \calH}$}
    \IF{$\mathtt{ beam.ys1[-1]==<\!\!eos\!\!>}$}
    \STATE $\mathtt{\calC.append(finish(beam))}$
    \ENDIF
    \ENDFOR
    \ENDWHILE
  \end{algorithmic}
\end{algorithm}

\section{Experiments}
\label{s:expt}
\vspace*{-1ex}

In the experiments, we largely adopt the acoustic modeling recipe based on transformers~\cite{Vaswan_17a} from Espnet~\cite{Watanab_18a}, as detailed in~\cite{Karita_19a}, for training the hybrid attention + CTC model~\cite{Watanab_17a}.\footnote{We mainly adapt Espnet's character BPE recipe for Switchboard.} 

We explore with two model architectures of different size. For the default, small architecture, the encoder shared by both attention and CTC consists of 2 convolutional layers that reduce the time and frequency dimension by a factor of 4, and 12 transformer layers, while the decoder consists of 6 transformer layers. Every encoder layer employs self-attention, and every decoder layer employs self-attention (to the previously decoded labels) followed by source-attention (to encoder outputs). All attention operations use 4 heads of 64 dimensions each, and the output of multi-head attention goes through a one-hidden-layer position-wise feed-forward network of 2048 ReLU units, before it is fed to the next layer. 
For the large architecture, we increase the number of transformer layers to 24 and 12 for the encoder and decoder respectively. The number of attention heads is increased to 6, yielding an attention dimension of 384. To improve generalization of the large architecture, we apply the technique of stochastic residual connections from~\cite{Pham_19a}. During training, we randomly skip the attention and feed-forward operations with a probability for each layer so the layer reduces to the identity mapping, and the layer dropout probability linearly increases with depth up to $p$.
The large architecture uses mostly the same hyperparameters tuned with the default architecture, and its results are denoted with ``stochastic layers ($p$)''.

We extract 80D fbank features plus 3D pitch features from audio 
(resampled to 16KHz) as inputs to acoustic model. We apply SpecAugment~\cite{Park_19a}, with ``max\_time\_warp'' set to 5 (frames), two frequency masks of widths up to 30 frequency bins, and two time masks of widths up to 40 frames during training.
A warmup schedule is applied to ADAM~\cite{KingmaBa15a} learning rate.
We average weight parameters of last 10 epochs to obtain the final model. A beam size of $20$ is used for decoding.

\begin{table}[t]
 \caption{Dev WERs (\%) of BPE systems with different number of units $k$ for WSJ. LM weights $(\alpha,\,\beta)$ are given in parenthesis.}
 \label{t:wsj-dev}
\vspace*{-2ex}
\centering
\begin{tabular}{@{}|@{\hspace*{0.012\linewidth}}c@{\hspace*{0.012\linewidth}}|@{\hspace*{0.012\linewidth}}c@{\hspace*{0.012\linewidth}}|@{\hspace*{0.012\linewidth}}c@{\hspace*{0.012\linewidth}}|@{\hspace*{0.012\linewidth}}c@{\hspace*{0.012\linewidth}}|@{\hspace*{0.012\linewidth}}c@{\hspace*{0.012\linewidth}}|@{\hspace*{0.012\linewidth}}c@{\hspace*{0.012\linewidth}}|@{\hspace*{0.012\linewidth}}c@{\hspace*{0.012\linewidth}}|@{}} 
\hline
Systems & $k\!=\!75$ & $100$ & $150$ & $250$ & $500$ & $1000$ \\
\hline
\caja{c}{c}{Char BPE\\[-.5ex]Subword ($\beta=0.8$)} &  9.6 & 9.8 & 10.3 & 10.9 & 11.2 & 12.0  \\ \hline
\caja{c}{c}{Char BPE\\[-.5ex]Multi-level ($0.6,\,1.0$)} &  7.4 & 7.5 & 8.3 & 9.0 & 9.1 & 10.1 \\ \hline
\caja{c}{c}{Phone BPE\\[-.5ex]Multi-level ($0.6,\,1.0$)} & 6.2 & 6.5 & 7.0 & 7.6 & 8.3 & 9.1 \\ \hline
  \end{tabular}
\vspace*{-1ex}
 \end{table}

\begin{table}[t]
 \caption{Test WERs (\%) on WSJ. $k=75$ for BPE systems.}
 \label{t:wsj-test}
\vspace*{-2ex}
\centering
\begin{tabular}{@{}|@{\hspace*{0.008\linewidth}}c@{\hspace*{0.008\linewidth}}|@{\hspace*{0.008\linewidth}}c@{\hspace*{0.008\linewidth}}|@{\hspace*{0.008\linewidth}}c@{\hspace*{0.008\linewidth}}|@{\hspace*{0.008\linewidth}}c@{\hspace*{0.008\linewidth}}|@{\hspace*{0.008\linewidth}}c@{\hspace*{0.008\linewidth}}|@{\hspace*{0.008\linewidth}}c@{\hspace*{0.008\linewidth}}|@{}} 
\hline
\caja{c}{c}{Unit/\\[-.5ex]LM}
&
\caja{c}{c}{Char\\[-.5ex]Word LM} & 
\caja{c}{c}{Char BPE\\[-.5ex]Subword} &
\caja{c}{c}{Char BPE\\[-.5ex]Multi-level} &
\caja{c}{c}{Phone BPE\\[-.5ex]Multi-level} & 
\caja{c}{c}{Joint\\[-.5ex]$\gamma\!\!=\!\!0.2$} \\
\hline
WER & 4.9 & 7.1 & 5.1 & 3.6 & 3.4 \\ \hline
\multicolumn{2}{@{}|@{\hspace*{0.008\linewidth}}c@{\hspace*{0.008\linewidth}}|@{\hspace*{0.008\linewidth}}}{Stoc. layers (0.5)} & 6.0 & 4.1 & 3.1 & 3.0 \\
\hline
  \end{tabular}
\vspace*{-2ex}
 \end{table}


\vspace*{-1ex}
\subsection{Results on Wall Street Journal (WSJ)}
\label{s:wsj}
\vspace*{-1ex}

We first verify the effectiveness of phone BPE systems on the WSJ corpora (
LDC93S6B and LDC94S13B). Partitions \emph{si284}/\emph{dev93}/\emph{eval92}  are used as the training/development/test set respectively. Mini-batch size is set to $16$ 
for $100$ training epochs.
Both subword and word LMs are trained on the WSJ language model training data, and the word LM has a vocabulary size of $65K$. To build the lexicon, 
we use \emph{cmudict}~\cite{cmudict} and apply the 
g2p model from~\cite{g2p-seq2seq} to words not in \emph{cmudict}.

We vary the number of subword units, denoted by $k$, for both character and phone BPE systems. Dev set WERs of these systems are given in Table~\ref{t:wsj-dev}. 
For character BPE systems, we use both subword RNNLM and multi-level RNNLM for decoding. We observe that small $k$ is clearly preferred by both systems, indicating that the WSJ training set (80 hours) probably has poor coverage for large set of BPE units; this observation is in line with that of~\cite{Xu_19a}.\footnote{We have trained a phone BPE system with $k=46$, where the BPEs include only the individual phones; this simulates a phone-based baseline and gives 6.3\% WER on the dev set.}
Furthermore, phone BPE systems consistently outperforms character BPE systems for all $k$ with multi-level LM. 
By investigating dev set learning curves for BPE systems with different $k$ (not shown due to space limit), we observe too large $k$ (e.g., $1000$) yields significantly worse losses, agreeing with the trend in WER. And for the same $k$, the phone BPE systems have consistently lower losses than character BPE systems, implying less confusion for the acoustic model.

We obtain test WERs of systems with $k=75$ in Table~\ref{t:wsj-test}. As a reference, we provide the WER of Espnet's character recipe trained with our setup (specAugment, same mini-batch size). With word RNNLM for decoding, it is not surprising that the character system, using $52$ character units, performs very similarly to the character BPE system with $k=75$ (4.9\% vs. 5.1\% test WER). The phone BPE system significantly outperforms character based systems, and joint decoding which emphasizes the phone BPE system (using $\gamma=0.2$) yields small further improvement. We note that the improvement from using phone-based subwords is on par with the one obtained from discriminative training for the character system achieved by~\cite{Baskar_19b}.
We then train the large architecture and give resulting WERs 
in Table~\ref{t:wsj-test} (last row). Stochastic transformer layers~\cite{Pham_19a} leads to improved generalization for large models.

\begin{table}[t]
 \caption{RT-03 WERs (\%) of phone-based BPE systems with different $k$ trained on SWBD (193K utterances). $\alpha\!=\!0$, $\beta\!=\!0.4$.}
 \label{t:swbd-dev}
\vspace*{-2ex}
\centering
\begin{tabular}{@{}|c|c|c|c|c|c|c|@{}} 
\hline
 & $k\!=\!50$ & $250$ & $350$ & $500$ & $1000$ & $2000$  \\
\hline
\caja{c}{c}{Phone BPE} & 15.2 & 14.7 & 14.7 & 14.5 & 14.4 & 15.5 \\ \hline
  \end{tabular}
\vspace*{-2ex}
 \end{table}

\begin{table}[t]
 \caption{WERs (\%) of BPE systems on eval2000 and RT-03.}
 \label{t:swbd-test}
\vspace*{-2ex}
\centering
\begin{tabular}{@{}|@{\hspace*{0.01\linewidth}}c|c@{\hspace*{0.01\linewidth}}|@{\hspace*{0.01\linewidth}}c@{\hspace*{0.01\linewidth}}|c@{\hspace*{0.01\linewidth}}|@{}} 
\hline
& \multicolumn{2}{c|}{eval2000} & \\ \cline{2-3}
\raisebox{1.5ex}{Modeling units} & SWBD & CALLHM & \raisebox{1.5ex}{RT03} \\
\hline
Char + sMBR~\cite{Cui_18a} & 12.0 & 23.1 & \\
Char BPE~\cite{Zeyer_18b} & 11.0 & 23.1 & \\
SentencePiece~\cite{Wang_19a} & 9.2 & 19.1 & \\
WordPiece~\cite{Park_19a} & 6.8 & 14.1 & \\
\hline


\textbf{Ours}: Stoc. layers (0.5), $\alpha\!\!=\!\!0$ &&&  \\
Char BPE (Subword, $\beta\!\!=\!\!0.2$) & 7.0 & 14.5 & 12.8 \\
Phone BPE (Multi, $\beta\!\!=\!\!0.4$) & 6.8 & 14.4 & 12.3 \\
Joint ($\beta\!\!=\!\!0.4$, $\gamma\!\!=\!\!0.4$) & 6.3 & 13.3 & 11.4 \\

\textbf{+ Fisher}: Stoc. layers (0.2) &&&  \\
Char BPE (Subword, $\beta\!\!=\!\!0.1$) & 5.1 & 9.5 & 8.5 \\
Phone BPE (Multi, $\beta\!\!=\!\!0.2$) & 5.5 & 10.0 & 9.1 \\
Joint ($\beta\!\!=\!\!0.2$, $\gamma\!\!=\!\!0.4$) & 4.9 & 9.5 &  \\
\hline                                                          
  \end{tabular}
\vspace*{-2ex}
 \end{table}

\subsection{Results on Switchboard (SWBD)}
\label{s:swbd}
\vspace*{-1ex}

We now experiment with the Switchboard corpus (
LDC97S62), with 300 hours of training data. 
In the first set of experiments, to explore the effect of $k$, we hold out 4K utterances from the full training set as development set, and set the mini-batch size to $256$ for training the small architecture on the remaining 193K utterances for 100 epochs. 
For decoding, the word LM has a vocabulary size of 31K and is trained on the SWBD training set transcription.
When $k=50$, the BPEs include only individual phones, 
simulating a phone-based baseline. 
We show the performance of phone BPE systems on RT-03 (LDC2007S10) in Table~\ref{t:swbd-dev}. Observe that SWBD prefers a much larger $k$ than WSJ, and the performance is stable for a large range of $k$.

We then add the 4K utterances back for training 
%
the large architecture, with minibatch size of 512 for 150 epochs. With PyTorch's DistributedDataParallel scheme, training takes 2 days with 8 Tesla V100 GPUs. We use a word RNNLM trained also on Fisher text for decoding, with a vocabulary size of 67K. 
Test WERs on \emph{eval2000} (LDC2002S09 and LDC2002T43) are provided in Table~\ref{t:swbd-test}, together with recent results obtained by attention-based models.\footnote{Trained on 193K utterances with the small architecture, the original Espnet recipe obtains 8.5\%/17.3\% WER on SWBD/CALLHM.}
The character BPE system no longer benefits from the word LM. With stochastic layers, the larger architecture significantly outperforms the small one. Our individual systems are on par with the large model from~\cite{Park_19a}, and joint decoding improves over known results in this setup.

Finally, we add the Fisher corpora (LDC2004T19 and LDC2005T19) to training, without tuning the architecture and most hyperparameters. 
Models converge after 60 epochs over the joint training set.
We tune LM weight for decoding, and with more acoustic training data, systems now prefer smaller $\beta$'s. Our phone BPE system performs similarly to previous best single system~\cite{Han_18a}, while the char BPE system, to our knowledge, sets the new state-of-the-art for single system with 5.1\%/9.5\% test WER. Joint decoding further improves the WER on the SWBD portion to 4.9\%, outperforming known SWBD results even by large ensemble of systems~\cite{Han_18a}.


\vspace*{-1ex}
\section{Conclusions}
\label{s:conclusion}
\vspace*{-1ex}

We have investigated the use of phone-based subwords in end-to-end ASR, proposed a novel multi-level language model for subword-based decoding, and a new beam search algorithm for joint decoding with phone- and character-based subword systems.
Experiments on two benchmark datasets show that phone-based BPE systems tend to achieve higher accuracy while maintaining the simplicity of end-to-end methods.
In the future, we can explore other types of subwords than BPE, and incorporate subword regularization~\cite{DrexlerGlass19a} which is shown to improve character-based subword systems.

\section{Acknowledgements}
Weiran Wang would like to thank Tong Niu for discussions on BPEs, Guangsen Wang for suggesting stochastic transformer layers, Shinji Watanabe for discussions on the Espnet recipe, and Karen Livescu for the reference~\cite{Xu_19a} (by Shinji Watanabe). 

\bibliographystyle{IEEEtran}
\bibliography{refs}
\end{document}